\documentclass[twocolumn]{aastex62}

\usepackage{gensymb}
\usepackage{booktabs}

\submitjournal{ApJ}

\shorttitle{HD 143006: Multiple rings, a misaligned inner disk, and a bright arc}
\shortauthors{P\'erez et al.}

\begin{document}

\title{The Disk Substructures at High Angular Resolution Project (DSHARP): \\ X. Multiple rings, a misaligned inner disk, and a bright arc in the disk around the T\,Tauri star HD\,143006}

\correspondingauthor{L. P\'erez}
\email{lperez@das.uchile.cl}

\author[0000-0002-1199-9564]{Laura M. P\'erez}
\affil{Departamento de Astronom\'ia, Universidad de Chile, Camino El Observatorio 1515, Las Condes, Santiago, Chile}

\author[0000-0002-7695-7605]{Myriam Benisty}
\affil{Departamento de Astronom\'ia, Universidad de Chile, Camino El Observatorio 1515, Las Condes, Santiago, Chile}
\affiliation{Unidad Mixta Internacional Franco-Chilena de Astronom\'{i}a, CNRS/INSU UMI 3386}
\affiliation{Univ. Grenoble Alpes, CNRS, IPAG, 38000 Grenoble, France.}

\author{Sean M. Andrews}
\affil{Harvard-Smithsonian Center for Astrophysics, 60 Garden Street, Cambridge, MA 02138, USA}

\author[0000-0002-0786-7307]{Andrea Isella}
\affiliation{Department of Physics and Astronomy, Rice University 6100 Main Street, MS-108, Houston, TX 77005, USA}

\author[0000-0002-7078-5910]{Cornelis P. Dullemond}
\affil{Zentrum f\"r Astronomie, Heidelberg University, Albert Ueberle Str. 2, 69120 Heidelberg, Germany}

\author[0000-0001-6947-6072]{Jane Huang}
\affil{Harvard-Smithsonian Center for Astrophysics, 60 Garden Street, Cambridge, MA 02138, USA}

\author{Nicol\'as T. Kurtovic}
\affil{Departamento de Astronom\'ia, Universidad de Chile, Camino El Observatorio 1515, Las Condes, Santiago, Chile}

\author[0000-0003-4784-3040]{Viviana V. Guzm\'an}
\affil{Joint ALMA Observatory, Avenida Alonso de C\'ordova 3107, Vitacura, Santiago, Chile}
\affil{Instituto de Astrof\'isica, Pontificia Universidad Cat\'olica de Chile, Av. Vicu\~na Mackenna 4860, 7820436 Macul, Santiago, Chile}

\author[0000-0003-3616-6822]{Zhaohuan Zhu}
\affiliation{Department of Physics and Astronomy, University of Nevada, Las Vegas, 4505 S. Maryland Pkwy, Las Vegas, NV, 89154, USA}

\author[0000-0002-1899-8783]{Tilman Birnstiel}
\affil{University Observatory, Faculty of Physics, Ludwig-Maximilians-Universit\"at M\"unchen, Scheinerstr. 1, 81679 Munich, Germany}

\author[0000-0002-8537-9114]{Shangjia Zhang}
\affiliation{Department of Physics and Astronomy, University of Nevada, Las Vegas, 4505 S. Maryland Pkwy, Las Vegas, NV, 89154, USA}

\author[0000-0003-2251-0602]{John M. Carpenter}
\affil{Joint ALMA Observatory, Avenida Alonso de C\'ordova 3107, Vitacura, Santiago, Chile}

\author[0000-0003-1526-7587]{David J. Wilner}
\affil{Harvard-Smithsonian Center for Astrophysics, 60 Garden Street, Cambridge, MA 02138, USA}

\author{Luca Ricci}
\affil{Department of Physics and Astronomy, California State University Northridge, 18111 Nordhoff Street, Northridge, CA 91130, USA}

\author[0000-0003-1172-3039]{Xue-Ning Bai}
\affil{Institute for Advanced Study and Tsinghua Center for Astrophysics, Tsinghua University, Beijing 100084, China}

\author[0000-0002-0786-7307]{Erik Weaver}
\affiliation{Department of Physics and Astronomy, Rice University 6100 Main Street, MS-108, Houston, TX 77005, USA}

\author{Karin I. \"Oberg}
\affil{Harvard-Smithsonian Center for Astrophysics, 60 Garden Street, Cambridge, MA 02138, USA}

\begin{abstract}
We present a detailed analysis of new ALMA observations of the disk around the T-Tauri star HD\,143006, which at 46\,mas (7.6\,au) resolution reveal new substructures in the 1.25\,mm continuum emission. The disk resolves into a series of concentric rings and gaps together with a bright arc exterior to the rings that resembles hydrodynamics simulations of a vortex, and a bridge-like feature connecting the two innermost rings.
Although our $^{12}$CO observations at similar spatial resolution do not show obvious substructure, they reveal an inner disk depleted of CO emission. 
From the continuum emission and the CO velocity field we find that the innermost ring has a higher inclination than the outermost rings and the arc. 
This is evidence for either a small ($\sim8\degree$) or moderate ($\sim41\degree$) misalignment between the inner and outer disk, depending on the specific orientation of the near/far sides of the inner/outer disk.
We compare the observed substructures in the ALMA observations with recent scattered light data from VLT/SPHERE of this object.
In particular, the location of narrow shadow lanes in the SPHERE image combined with pressure scale height estimates, favor a large misalignment of about $41\degree$.
We discuss our findings in the context of a dust-trapping vortex, planet-carved gaps, and a misaligned inner disk due to the presence of an inclined companion to HD\,143006.

\end{abstract}

\keywords{ALMA --- protoplanetary disks}

\section{Introduction} \label{sec:intro}

High resolution observations of protoplanetary disks have shown that most, if not all, disks host substructures in the form of spiral arms, rings, or azimuthally asymmetric features. These features are imprints of the processes of disk evolution and planet formation, and can be observed from optical \citep[in scattered light, e.g.][]{avenhaus2018} to millimeter wavelengths \citep[e.g.][]{rings,spirals}. In the dust continuum, the most common substructures are multiple bright rings and dark gaps, possibly tracing over-densities and depletion of dust grains, respectively. Rings and gaps are a natural outcome of dynamical interactions with planets \citep[e.g.,][]{Paardekooper2004,crida2007,zhu2011}, but  could also be tracing ice lines \citep[e.g.,][]{zhang2015,okuzumi2016}, or be a result of certain magneto-hydrodynamical processes \citep[e.g.,][]{johansen2009,dittrich2013,bai2014,simon2014,bethune2017,suriano2018}.

These substructures could play a very important role in the process of planet formation, and in particular, in the first steps towards the growth of planetesimals. Alternatively, substructures may exhibit what is the outcome of planet-disk interaction, and serve as an important diagnostics of ongoing planet formation.
Since dust particles experience aerodynamic drag from the gas, in a smooth disk they would rapidly lose angular momentum and drift towards the star before they can grow to the planetesimal size \citep{brauer2007, birnstiel2010}. However, dust grains can be maintained in localized particle traps (pressure maxima) which allow them to efficiently grow \citep{pinilla2012b}. Observationally, these traps appear as a suite of bright and dark rings in the dust continuum \citep{pinilla2012a}, reminiscent of continuum observations of multiple ringed systems \citep[e.g.,][]{ALMA}. In addition to concentric rings, azimuthal asymmetries have also been observed, in particular in protoplanetary disks with dust-depleted large cavities \citep[transition disks; e.g.,][]{casassus2013, isella2013, dong2018b}. They are interpreted as azimuthal dust trapping \citep{birnstiel2013,Lyra}, possibly in a vortex due to the Rossby wave instability \citep[e.g.,][]{li2000,meheut2012, zhu2014}. Such an instability can be generated at the edge of a gap created by a massive planet \citep[e.g.,][]{regaly2012,ataiee2013,zhu2016}, or at the edge of a dead zone \citep[e.g.,][]{kretke2007,flock2017}. 

Planet-disk interactions can also leave imprints on the disk gas kinematics \citep{seba2015} and observations of CO isotopologues have revealed non Keplerian velocities in various disks. Recently, \citet{pinte2018} found a local deformation of the CO disk velocity field of the intermediate-mass young star HD\,163296, consistent with the spiral wave induced by a  2\,M$_{\rm{Jup}}$ planet located at 260\,au. \citet{teague2018} measured, in the same disk, local pressure gradients consistent with gaps carved by a 1\,M$_{\rm{Jup}}$ planet at 83\,au and 137\,au.  Other observations of non-Keplerian velocities were explained by the presence of radial flows \citep{rosenfeld2014}, of a warp \citep{rosenfeld2012,casassus2015,walsh2017} or of free-falling gas connecting a strongly misaligned inner disk to the outer disk  \citep{loomis2017}. 
A misalignment between the inner and outer disk could be induced by a massive companion on an inclined orbit with respect to the plane of the disk \citep[e.g.][]{bitsch2013,owen2017}, or alternatively, by a misaligned stellar magnetic field \citep{bouvier2007} in which case the star should be periodically obscured \citep[periodic dippers, see e.g.,][]{cody2014}. In the presence of a strongly misaligned inner disk, the outer disk illumination is drastically affected, as evidenced by shadows observed in scattered light images that trace the surface of the disk  \citep[e.g.][]{marino2015,benisty2017,benisty2018,casassus2018,pinilla2018}. 

The focus of this paper is the T Tauri star HD\,143006, a G-type star \citep[$1.8^{+0.2}_{-0.3}$ M$_{\odot}$, 4-12 Myr old,][]{preibisch2002,pecaut2012,garufi2018,survey}, located in Upper Sco at a distance of $165\pm5$\,pc \citep{Gaia2018}. Sub-millimeter continuum observations of the disk at moderate spatial resolution revealed an azimuthal asymmetry and a marginally-resolved cavity \citep{barenfeld2016}, although it is not classified as a transition disk from its spectral energy distribution. A strong near-infrared excess and typical mass accretion rate \citep[$\sim2\times10^{-8}\,M_{\odot}$\,yr$^{-1}$;][]{rigliaco2015} indicate that the innermost regions retain dust and gas. Recent scattered light observations of HD\,143006 show multiple brightness asymmetries: a shadow over half of the disk (the West side) and two additional narrow shadow lanes, both suggestive of a moderately-misaligned inner disk \citep[][and \S\ref{sec:SPHERE}]{benisty2018}. 

In the following, we present new observations of HD\,143006 obtained with ALMA as part of DSHARP in \S\ref{sec:obs}, we then present a characterization of its substructures in \S\ref{sec:results}, that we further discuss in \S\ref{sec:discussion}, and the conclusions of our work in \S\ref{sec:conclusions}.

\begin{figure*}[t!]
\includegraphics[width=1.\textwidth]{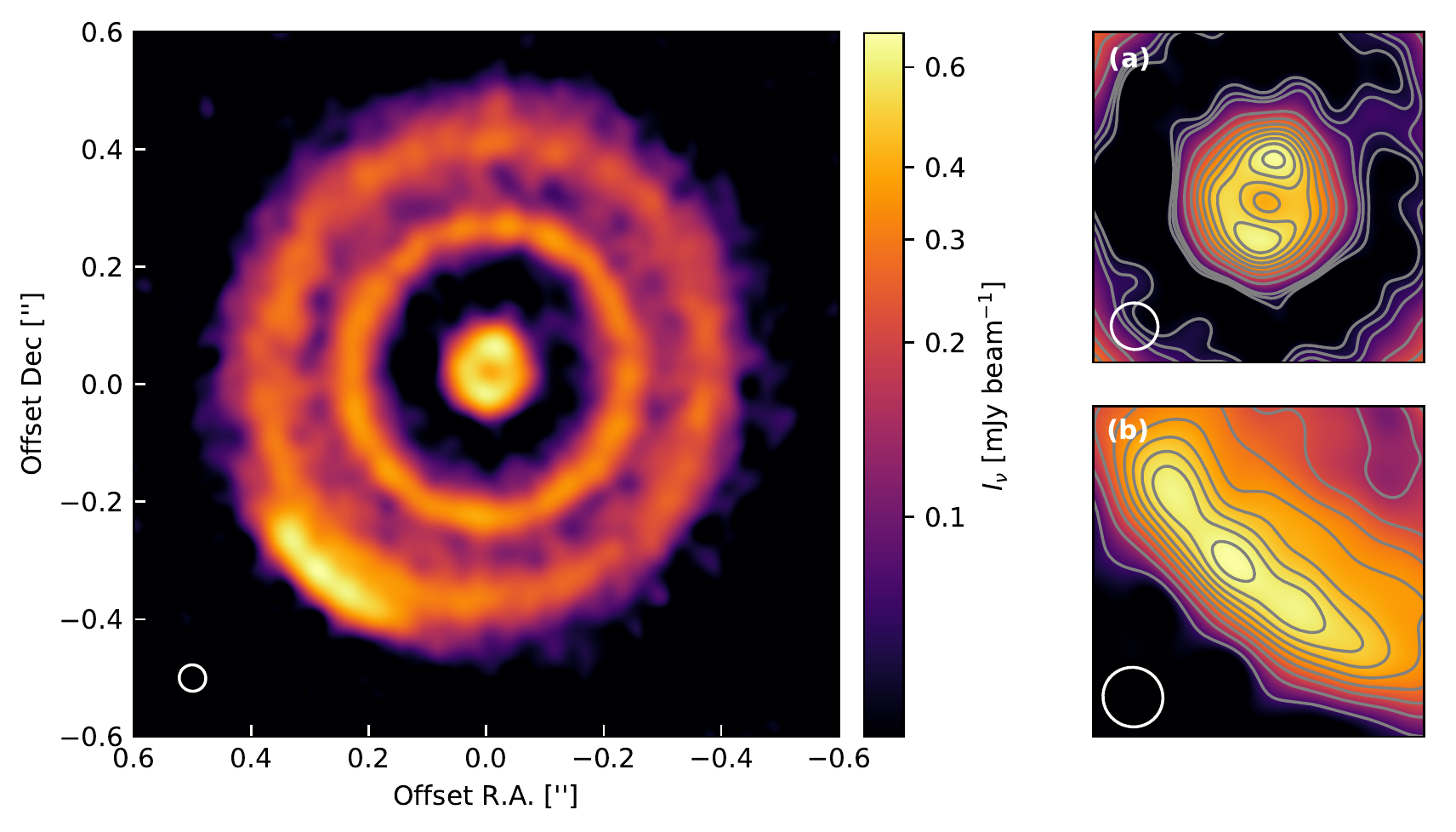}
\caption{{\emph Left:} ALMA observations at 1.25\,mm of the dust continuum emission from the disk around the young star HD\,143006. \emph{Right:} panels show a zoom to (a) the inner disk within 0.32\arcsec{}, and (b) the arc substructure within 0.25\arcsec{}. In panel (a), the contours start at 3, 4, 5$\sigma$, then are spaced by $5\sigma$, while in panel (b) contours start and are spaced by $5\sigma$, where $\sigma$ is the RMS noise of the image ($\sigma=14.3$ $\mu$Jy beam$^{-1}$).
Ellipses on the bottom-left corner of each panel indicate the beam size of $46\times45$ mas ($7.6\times7.4$ au). \label{fig:cont} }
\end{figure*}

\section{Observations} \label{sec:obs}

Dust continuum emission at 1.25\,mm and $^{12}$CO emission in the $J=2-1$ transition were observed with long and short baselines in the ALMA Large Program 2016.1.00484.L. Additional short baselines observations (from project 2015.1.00964.S, P.I.\ \"Oberg) in the same dust and gas tracers were combined to increase uv-coverage on the short-spacings. Details on the calibration of visibilities can be found in \cite{survey}. Observations were centered to R.A.\,(J2000) = 15h\,58m\,36.90s, Dec\,(J2000) = $-22$d\,57m\,15.603s, which are the coordinates of the phase center of the last execution of the long baseline observations. The continuum images presented here used the same imaging parameters as listed in Table 4 of \citet{survey}, while the CO images were obtained using a robust parameter of 0.8 with a uv-tapering of $20\;$mas, which was applied to the CO visibilities to enhance extended emission from the disk. The RMS noise in the continuum image is 14.3\,$\mu$Jy\,beam$^{-1}$ for a $46\times 45$\,mas beam with position angle (PA) of $52.1\degree$. In the CO spectral cube we measure an RMS noise of 1.04\,mJy\,beam$^{-1}$\,km\,s$^{-1}$, for a beam size of $66\times49$\,mas, PA of $83.6\degree$, and channel width of 0.32\,km\,s$^{-1}$. The continuum map is presented in Figure \ref{fig:cont}, and the full set of CO channel maps in Figure 5.9 of \citet{survey}.

\begin{figure*}[t!]
\includegraphics[width=1.\textwidth]{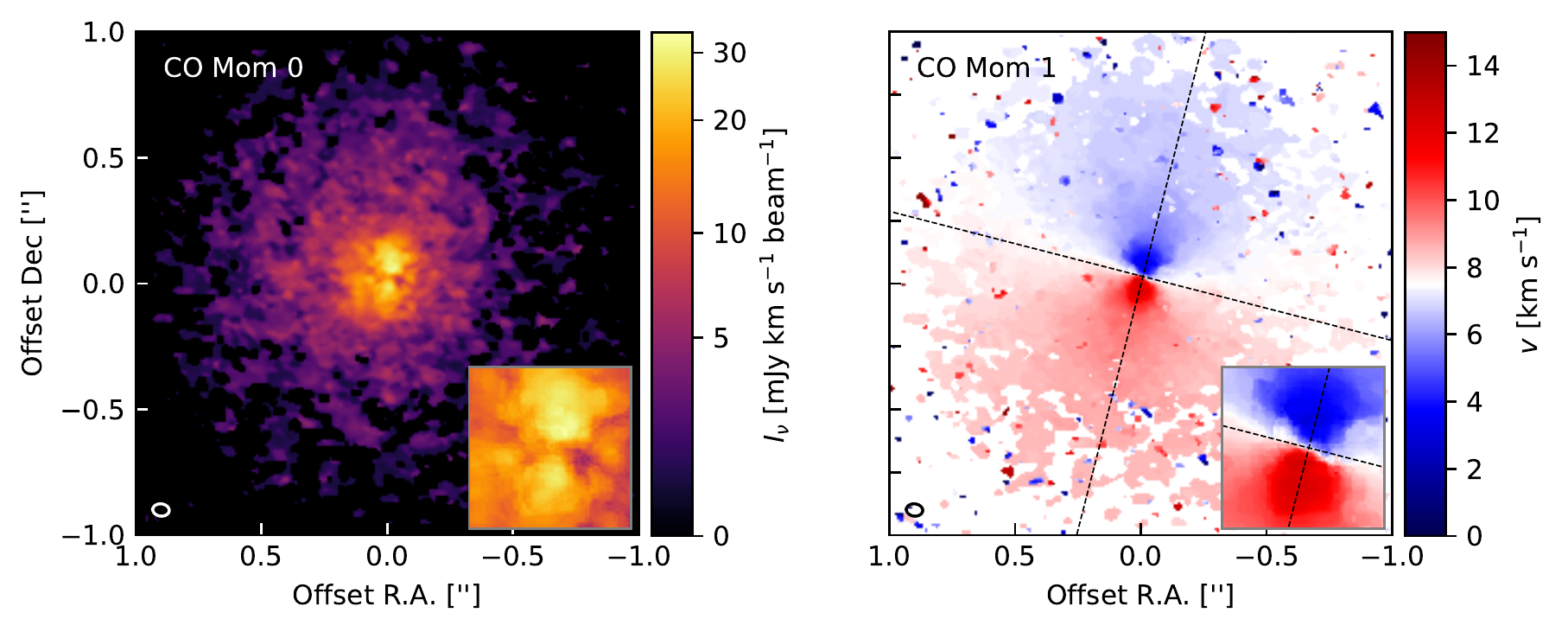}
\caption{ALMA observations of the $^{12}$CO$(J=2-1)$ line from the disk around the young star HD\,143006. Note that the field of view is $2''\times2''$, larger than in Figure 1. Left panel: map of the integrated intensity of the CO line, right panel: map of the intensity-weighted velocity of the CO line. Insets on each panel show a zoom to the inner disk region within 0.25$''$. The \emph{global} PA of the disk, as constrained by the best-fit model in \S3.3 to be $167\degree$, is indicated by dashed lines. Ellipses on the bottom-left corner of each panel indicate beam size of $66\times49$ mas ($10.9\times8.1$ au). \label{fig:contCO} }
\end{figure*}

\section{Results} \label{sec:results}

\subsection{Dust Continuum Emission}
\label{sec:dust}

The continuum emission from HD\,143006, presented in Figure \ref{fig:cont}, is observed close to face-on: the disk resolves into an inner depleted region, three bright rings at roughly $0.05''$ (8\,au), $0.24''$ (40\,au), $0.39''$ (64\,au) from the disk center, two dark annuli at roughly $0.13''$ (22\,au) and $0.31''$ (51\,au), and a bright south-east asymmetric feature just outside of the outermost ring at $0.45''$ (74\,au) from the disk center. This prominent arc is located at a position angle (PA, defined from North toward East) of $\sim140\degree$ and has an azimuthal extent of $\sim30\degree$.
From now on, we follow the convention of \citet{rings} and label the dark/bright annular features with a letter ``D''/``B'' followed by their radial location in au. Then, the innermost bright ring is labeled B8, the second bright ring as B40 and the outermost bright ring as B64. The dark annuli between B8 and B40, and between B40 and B64 are labeled D22 and D51, respectively. 

The radial profile of the continuum emission presented in \citet{rings} indicates that B8 and B40 have a contrast of $\sim42$ and $\sim24$ with respect to the D22; these are one of the highest contrast ring-features of the DSHARP sample. 
However, these rings are not completely smooth: B8 has two peaks along a PA of $\sim165\degree$ (see panel (a) on Fig.\ref{fig:cont}), but this is due to its inclination w.r.t.\ our line of sight (see \S3.4). The outermost rings, B40 and B64, both peak close to a PA of $180\degree$ (better seen in the residual maps of \S\ref{sec:model}). 
The bright asymmetric feature outside of B64 (see panel (b) on Fig.\ref{fig:cont}) is also uneven in the azimuthal direction: the arc resolves into three peaks along its azimuthal extent. The brightest peak at the center of the arc has a signal-to-noise ratio (SNR) of 48, while the flanking peaks have a SNR of 44 and are separated by $\sim10$\,au (60\,mas) from the brightest peak of the arc.
Finally, the dark annuli are not completely empty, in particular, inside D22 there is a bridge-like emission feature, with a SNR of 4.9$\sigma$, connecting the B8 and B40 rings (see emission between the rings at a PA of $\sim 300\degree$ in panel (a) of Figure \ref{fig:cont}).

\subsection{Gas emission traced by CO}
\label{sec:gas}

The overall emission from the gaseous component of the HD 143006 disk, as traced by the CO line, is presented in the middle and right panels of Figure \ref{fig:contCO} and extends out to $\sim1''$ in radius, twice larger in radial extent than the dust continuum emission. CO emission is detected above 3$\sigma$ from $-0.12$ km s$^{-1}$ to $15.2$ km s$^{-1}$ in the LSRK reference frame, with a systemic velocity of $\sim7.7$ km s$^{-1}$.
The integrated intensity of CO, also known as moment 0 map (left panel, Figure~\ref{fig:contCO}) is not centrally peaked. Since the continuum emission appears to be optically thin \citep[$\tau<0.2$,][]{rings}, the lack of CO emission in the inner disk cannot be fully explained by continuum subtraction of optically-thick dust emission, and rather indicates that some gas depletion occurs in the inner disk.

The intensity-weighted velocity of CO, also known as moment 1 map (right panel, Figure~\ref{fig:contCO}), shows a clear rotation pattern for a nearly face-on disk with a PA of $\sim165\degree$ (note that the moment\,0 map has two maxima along a similar PA). However, the disk rotation is not perfectly described by a single geometry, as can be seen on the inset of Figure~\ref{fig:contCO}, right panel, where the velocity field in the inner disk appears different from the outer disk. This difference will be quantified in the next sections.

Given the low inclination of this system and the low SNR of the spectral cube (peak SNR of $\sim 11$), the front and back sides of the CO emitting layers are not clearly separated, a feature observed in other disks with higher line-of-sight inclinations \citep[e.g.,][]{rosenfeld2013,isellaLP}. Nevertheless, of the two possibilities for the absolute disk geometry, we suggest that the West side of the disk is the one closest to us: at every channel, the West side of the disk appears shorter when projected onto the rotation axis of the disk  than the East side \citep[see Figure 5.9 in][]{survey}

Finally, a close inspection of the channel maps revealed that redshifted emission close to the systemic velocity has a deviation or ``kink'' from the Keplerian iso-velocity curves that is not present in the blue-shifted channels. This kink is marked by an arrow on the bottom panels (redshifted channels) of Figure \ref{fig:kink} (see Appendix) and appears at a PA of between $\sim80\degree-120\degree$.

\begin{figure*}[ht!]
\centering
\includegraphics[width=\textwidth]{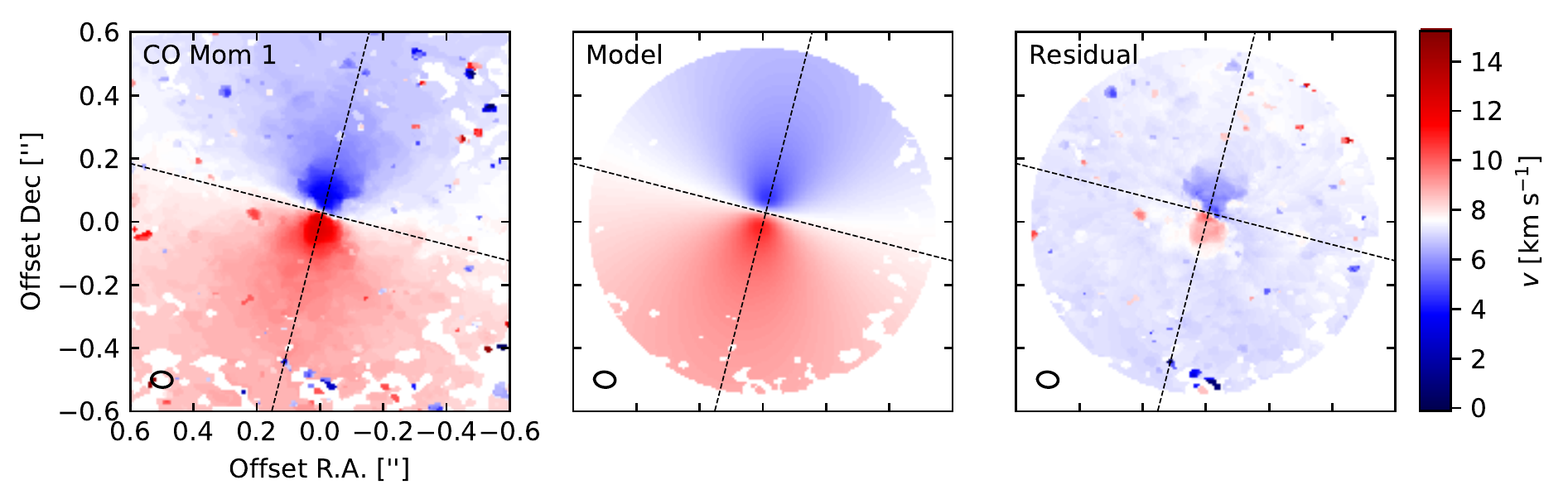}
\caption{Comparison between the intensity-weighted velocity field around HD\,143006 (right), the best-fit model for a Keplerian razor-thin disk (middle), and the residual velocity field map obtained from subtracting the best-fit from the observations (left). The \emph{global} PA of the disk, as constrained by the best-fit model to be $167\degree$, is indicated by dashed lines.  \label{fig:mom1}}
\end{figure*}

\subsection{Modeling CO kinematics} \label{sec:COmodel}

We model the Keplerian velocity field in HD\,143006  using an analytical model of a razor-thin disk in Keplerian motion around a 1.8\,M$_{\odot}$ star, in order to constrain the \emph{global} geometry of the disk. The model parameters are the inclination ($i$), position angle (PA), systemic velocity ($v_{sys}$), as well as the right ascension and declination offsets ($x_0,y_0$) for the center of rotation. We fix the inner and outer radius of the disk to 0.3 and 250\,au, respectively.
For a given set of parameters we generate a velocity map that is convolved with the beam of our CO observations, which we then compare to the intensity-weighted velocity of CO (right panel of Figure~\ref{fig:contCO}) pixel by pixel.  We perform our fit using an affine invariant MCMC sampler \citep[$emcee$,][]{emcee}, and sample the parameters with uniform priors in the following range: $i$ between [$0\degree, 90\degree$], PA between [$0\degree, 360\degree$], $v_{sys}$ between [5 km s$^{-1}$, 10 km s$^{-1}$], and $x_0,y_0$ between
[-0.2\arcsec{}, 0.2\arcsec{}]. The parameter space is explored with 80 walkers and for 4000 steps. From the last 2000 steps we compute the best-fit values, which correspond to the 50$^{th}$ percentile of the samples in the marginalized distributions, while the error bars correspond to the 16$^{th}$ and 84$^{th}$ percentiles. The best-fit parameters can be found on Table\,\ref{tab:COmodel}, while Figure \ref{fig:mom1} presents the velocity field of the observations, best-fit model, and residual (computed by subtracting the model from the observations). Although the best-fit model reproduces the velocity field of the outer disk, inwards of $\sim0.15''$ there are significant residuals that suggest gas moving at higher speeds than those in the model. 

Note that if we separately fit the inner and outer disk kinematics, by masking the velocity field in Figure~\ref{fig:mom1} inside and outside of $0.15''$ (which corresponds to the mip-point separation between B8 and B40), we find two different disk geometries. First, a fit for the geometry of the outer disk while excluding the inner disk (i.e.\ masking those pixels inside of $0.15''$), results in a PA consistent with the global fit in Table~\ref{tab:COmodel} but a slightly more face-on outer disk with an inclination of $14.59\degree\pm0.02\degree$.
On the other hand, when we only fit for the disk kinematics inside of $0.15''$ (i.e. masking those pixels outside of $0.15''$), we find $i = 22.84\degree \pm 0.05\degree$ and PA~$= 165.6\degree\pm0.1\degree$ for the inner disk. Thus, we constrain a difference between the inner and outer disk geometry, in particular, the inclination of the disk changes from the inner to the outer disk (see Figure~\ref{fig:restrictedmom1} on the Appendix, for the best-fit model and residuals when the inner or outer disk are masked).

\begin{table}[!t]
\caption{Best-fit model to the intensity-weighted velocity field around HD\,143006} \label{tab:COmodel}
\begin{center}
\begin{tabular}{ l l   r r }
\toprule
\multicolumn{2}{l}{{\bf Parameter}} & & \multicolumn{1}{c}{\emph{Best-fit value}} \\
\cmidrule{1-4}
    $x_0$ 	& (mas) 		& & $-6 \pm 3$  \\
    $y_0$ 	& (mas) 		& & $29\pm 2$  \\
    $i$ 	& ($^\circ$)	& & $16.2\degree\pm0.3\degree$  \\
    PA	 	& ($^\circ$)	& & $167\degree\pm1\degree$  \\
    $v_{sys}$ & km s$^{-1}$	& &	$7.71\pm0.02$  \\
\bottomrule
\end{tabular}
\end{center}
\end{table}%

\begin{table*}[!t]
\caption{Best-fit model to the dust continuum emission in HD\,143006} \label{tab:dustmodel}
\begin{center}
\begin{tabular}{ l c   r r    r r    r r }
\multicolumn{2}{l}{{\bf Disk Geometry}} & & \multicolumn{1}{c}{\emph{Inner Disk}} &  & \multicolumn{3}{c}{\emph{Outer Disk}}	\\
\cmidrule{1-2}
\cmidrule{4-4}
\cmidrule{6-8}
$i$ 	& ($^\circ$)& & $24.1\pm1.0$	& & \multicolumn{3}{c}{$17.02\pm0.14$} \\
$PA$	& ($^\circ$)& & $164.3\pm2.4$	& & \multicolumn{3}{c}{$176.2\pm0.6$} \\
$x_0$	& (mas)		& & $-4.4\pm0.2$			& & \multicolumn{3}{c}{$-1.0\pm0.2$} \\
$y_0$ 	& (mas)		& & $23.1\pm0.2$			& & \multicolumn{3}{c}{$28.9\pm0.2$} \\
\cmidrule{1-8}
\multicolumn{2}{l}{{\bf Ring or Arc Parameters}} & &\multicolumn{1}{c}{B8} & & \multicolumn{1}{c}{B40} & \multicolumn{1}{c}{B64} & \multicolumn{1}{c}{Arc}	\\
\cmidrule{1-2}
\cmidrule{4-4}
\cmidrule{6-8}
$I_R$ or $I_A$				& ($\mu$Jy)	& & $1.54\pm0.02$	& & $0.64\pm0.01$	& $0.410\pm0.001$	    & $1.09\pm0.01$  \\
$r_R$ or $r_A$				& (au)		& & $7.67\pm0.04$	& & $39.95\pm0.03$	& $63.6\pm0.1$		& $74.2\pm0.1$ \\	
$\sigma_R$ or $\sigma_{r,A}$& (au)		& & $2.54\pm0.04$	& & $4.2\pm0.1$		& $9.4\pm0.1$		& $4.6\pm0.1$ \\
$\sigma_{\theta,A}$			& (au)		& & - 			& &	-					& -					 & $21.1\pm0.1$ \\ 
$\theta_A$ 					&($^\circ$)	& & -			& & -				 	& -					 & $141.4\pm0.1$ \\ 
\cmidrule{1-8}
\end{tabular}
\end{center}
\end{table*}%

\subsection{Modeling of continuum emission} \label{sec:model}

To constrain the observed substructure in this disk, we generate simple morphological models that describe the emission for the observed rings and the south-east arc. These models are constructed by combining the emission of three radially-Gaussian rings (that describe the B8, B40 and B64 features) and a two-dimensional Gaussian, in both the radial and azimuthal direction (that describes the outermost arc feature).

For every ring, its intensity is given by:
\begin{equation}
I(r) = I_R \; e^{-(r-r_R)^2/2\sigma_R^2},
\end{equation}
where $I_R$ is the peak intensity at radius $r_R$,  $\sigma_R$ is the ring width, and the ring is assumed to be symmetric along the azimuthal direction. For the asymmetric arc feature, its intensity is given by:
\begin{equation}
I(r,\theta) = I_A \; e^{-(r-r_A)^2/2\sigma_{r,A}^2} \; e^{-(\theta-\theta_A)^2/2\sigma_{\theta,A}^2}, 
\end{equation}
where $I_A$ is the peak intensity at radius $r_A$ and azimuthal location $\theta_A$, and $\sigma_{r,A}$, $\sigma_{\theta,A}$ are the width of the 2D-Gaussian in the radial and azimuthal direction, respectively.  The model adopted for the arc corresponds to the distribution of material in a steady-state vortex \citep{Lyra}.

\begin{figure*}[!t]
\includegraphics[width=\textwidth]{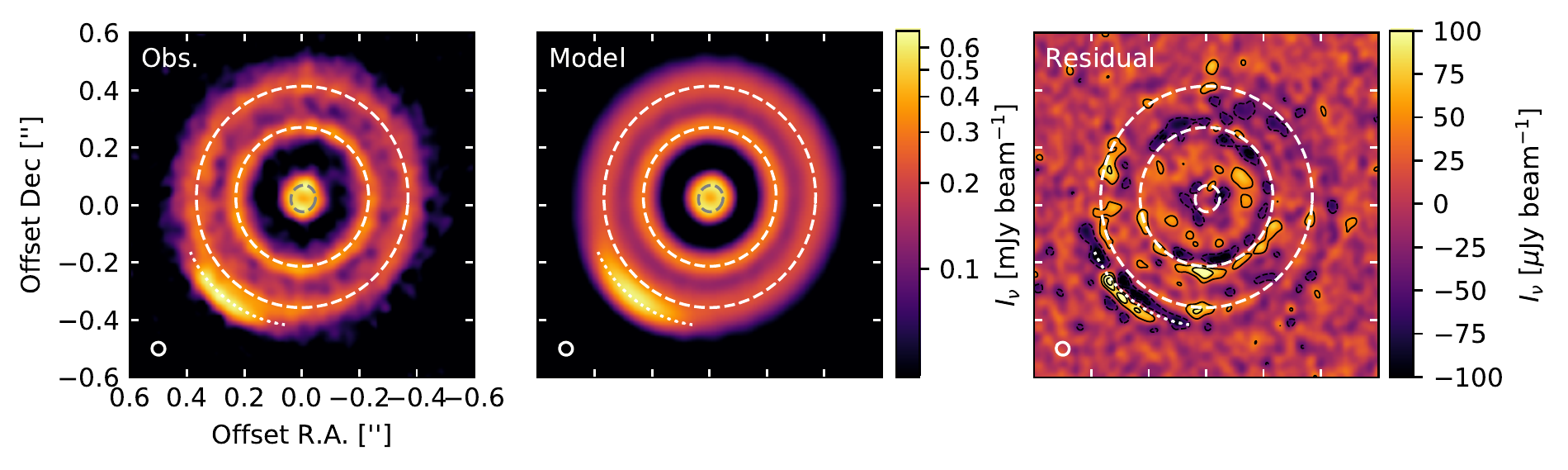}
\caption{Comparison between the 1.25\,mm observations from HD\,143006 (left) with the image from the best-fit rings+arc morphological model (middle) and the residual map from subtracting the best-fit model from the observations (right). The location and geometry of the features, as constrained by the best-fit model, are indicated by dashed lines for B8, B40, and B64, and a dotted line for the arc. Ellipses on the bottom-left corner of each panel indicate the beam size, contours on the residual map start at $3\sigma$ and are spaced by $\pm3\sigma$, where $\sigma$ is the RMS noise of the dust continuum image. \label{fig:modelCont}}
\end{figure*}

Additional parameters of importance are any offset ($x_{0},y_{0}$) that the rings may have from the phase center of the observations defined in \S2, and the inclination ($i$) of each feature along a particular position angle (PA) on the sky. 
Given the evidence for a misalignment between the inner disk and outer disk presented in \citet{benisty2018}, as well as the evidence from modeling its velocity field (\S3.3), we assume that the features found in the outer disk (i.e. B40, B64, and the arc) share the same center, inclination, and PA, while the innermost ring (B8) may have a different geometry due to the misalignment (i.e. different center, inclination, and PA). 
This results in a total of 22 free parameters that are constrained with over $34$\,million visibilities.
For a given set of parameters we produce a model image of the disk that is Fourier-transformed and sampled at the same locations in the uv-plane as the observed visibilities; for this we use the publicly available code Galario \citep{Galario}. To sample the PDF of our model parameters, we use the MCMC sampler $emcee$ \citep[][]{emcee} with 120 walkers that we ran for 25000 steps. Convergence was checked by measuring the autocorrelation time, which was under 3000 iterations. For each parameter we compute the posterior PDF by marginalizing over all but the parameter of interest over the last 15000 steps. 

The best-fit values of the continuum emission model, chosen as the 50$^{th}$ percentile of the PDF, as well as the 1-$\sigma$ uncertainty of each parameter from the 16$^{th}$ and 84$^{th}$ percentile of the PDF, are shown in Table \ref{tab:dustmodel}. From the best-fit we construct model visibilities (sampled at the same locations in uv-space as the observed visibilities) and we compute residual visibilities by subtracting the model visibilities from the observations. Images of the best-fit model and residual, obtained with the same imaging parameters as the observations (see \S2), are presented in Figure \ref{fig:modelCont}. 

The morphological models employed here can reproduce the observed emission reasonably well, as can be seen in the residual map of Figure \ref{fig:modelCont} where leftover emission is minimal and the largest residual is at $\pm9\sigma$. 
\citet{rings} find the rings locations directly on the image, resulting in differences of only 2-3\% for the outermost, well-resolved, rings and a 30\% difference in the radial location of B8, which due to its small angular size will have much less independent measurements on the image.
The radial width of the rings increases as a function of distance from the star: our model constrains radial extents of 6, 10, and 22\,au (in FWHM) for B8, B40, and B64. When fitting width of each ring directly on the image, \citet{dullemond2018} find a similar but narrower ring widths for B40 and B64, 7\% and 30\% narrower, respectively. 

As the rings are not completely symmetric in the azimuthal direction, for both B40 and B64 the radially-Gaussian rings cannot reproduce the excess of emission seen in the South of each ring. 
For the south-east arc, our model constrains a narrow extent in the radial direction with a factor of $\sim4$ wider extent in the azimuthal direction, similar to that observed in other disks with arcs observed at lower angular resolution \citep[e.g.][]{Perez2014}. There are significant residuals at the arc location, in particular, the 2D-Gaussian prescription cannot properly describe the multiple peaks observed along the arc (Fig.\,\ref{fig:cont}, panel (b)), which appear also in the residual map at the $\sim6-9\sigma$ level.

\newpage

\section{Discussion} \label{sec:discussion}

\subsection{Substructures in the ALMA images}

The distribution of larger solids, as traced by the ALMA dust continuum observations, reveals a wealth of substructure in the HD\,143006 disk (\S3.4), while the distribution of CO emission exhibits little substructure (only in the innermost disk regions, \S3.3). This difference may arise from the fact that the 1.25\,mm dust  emission is optically thin throughout the disk \citep{rings}, while the observed CO emission is optically thick and traces the surface layers of the disk.\\

\textbf{Bright rings and dark annuli.} 
For the dark annuli D22 and D51, there is not a corresponding decrement of CO emission in either the channel maps \citep[Figure 5.9][]{survey} or the integrated CO emission map (Figure~\ref{fig:contCO}, right panel). 
Assuming that the dust depletion observed in the outermost dark annulus originates from dynamical clearing by objects embedded in the disk, \citet{Shangjia2018} infer masses of $\sim10-20M_{Jup}$ and $\sim0.2-0.3M_{Jup}$, for planets inside D22 and D51, respectively.

\begin{figure*}[!t]
\includegraphics[width=\textwidth]{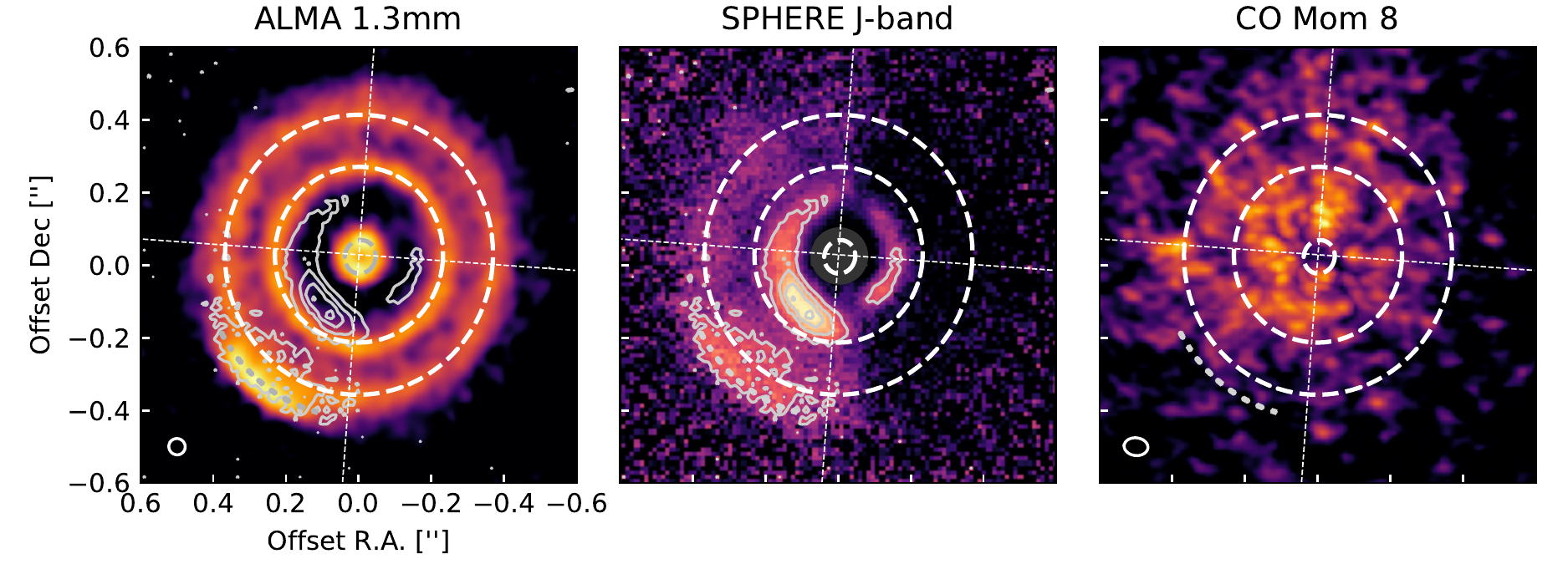}
\caption{Comparison between the ALMA continuum emission (left),  the SPHERE polarized intensity $J$-band observations (middle) and the $^{12}$CO peak emission map (right, also known as moment 8 map). The thin dashed lines go along and perpendicular to the PA of the outer disk, while the location and geometry of the features, as constrained by the best-fit model, are indicated by dashed lines for B8, B40, and B64, and a dotted line for the arc.
Contours are drawn at 25, 50, 75, and 95\% of the peak value of the SPHERE image on the middle panel. Ellipses on the bottom-left corner indicate the beam size for the ALMA observations. \label{fig:SPHERE}}
\end{figure*}

On the other hand, there is a decrement of both CO and dust emission inside B8. In particular, CO emission does not appear to be centrally-peaked, either on the integrated emission map (moment 0 map, left panel of Figure~\ref{fig:contCO}) or in the peak emission map (moment 8 map, right panel of Figure~\ref{fig:SPHERE}). 
Although both these maps are susceptible to beam dilution effects \citep{weaver2018}, which may cause an ``false'' inner cavity in the gas, the lack of CO emission inside B8 is also observed in the channel maps away from the systemic velocity \citep[see appendix Figure~\ref{fig:kink}, and Figure 5.9 in][]{survey}, meaning that there is some depletion of the CO column density inside of $\sim$15\,au ($\sim90$\,mas). We note that the lack of gas emission inside B8 cannot be explained by absorption from optically thick dust in the inner disk, since we also observe a dust emission deficit inside B8 and the emission from dust is optically thin throughout \citep{rings}.
Most likely, the depletion of CO inside B8 is quite large, for example, in transition disk cavities with centrally-peaked CO emission, the depletion has been measured to be more than an order of magnitude \citep{Marel2015}, and in the case of HD\,143006 we do not observe a centrally peaked CO. 

A possible origin for the depleted inner disk is photoevaporation, however, the star is accreting at a moderate rate \citep[$2\times10^{-8} M_{\sun}$ yr$^{-1}$,][]{rigliaco2015} and the near infrared excess points to a dust-rich inner disk, making this possibility less likely. 
A perturber inside B8 could deplete both gas and dust, and even misalign the inner disk, a possibility that will be discussed in \S4.3.

\textbf{Bright arc in the south-east.} Vortices are regions of higher gas pressure that can very efficiently trap solids \citep[e.g.,][]{barge1995,klahr1997,baruteau2016}. We model the arc emission with a prescription for the distribution of dust in a vortex that has reached steady-state \citep{Lyra}, and which should be smooth in the radial and azimuthal direction. However, the prominent arc in the 1.25\,mm continuum images has further substructure: three separate peaks that are unresolved in the radial direction (see residual map in Figure~\ref{fig:modelCont}), with a separation of $\sim10$\,au from the central peak. 
During their lifetime, vortices may never reach a stable equilibrium. For example, simulations have shown that when the back-reaction of the dust onto the gas is taken into account, vortices may trap particles not in a single but in multiple structures, and it is this dust feedback that would eventually lead to the destruction of the vortex \citep[see, e.g.,][]{fu2014}. Thus, it is not surprising that the observed arc has internal substructure.

The radial and azimuthal extent of the arc constrained by our model is about 5 by 21\,au. 
From the midplane temperature profile ($T_{mid}$) assumed in \citet{rings}, we estimate that at the vortex location ($\sim74$\,au), the pressure scale height corresponds to $H_p \sim 5$\,au (assuming a local sound speed of $c_s = (k_B T_{mid} / \mu m_p)^{0.5}$, and $\mu=2.3$).
Thus, the vortex is radially as wide as $H_P$ and it extends azimuthally over a few pressure scale heights, still consistent with the vortex scenario.
In particular, the $\sim21$\,au azimuthal extent is consistent with that of vortices formed at the edge of a gap/cavity, as it has a much smaller width than the $\pi$-wide vortices expected due to instabilities at the dead zone \citep{regaly2017}. However, the arc is outside of the millimeter dust disk, rather than at the edge of a dark annulus, and it is radially separated from B64 by $\sim 10$\,au. Such a configuration is similar to that of the disk in HD~135344B, which resolves into a narrow ring with an asymmetric arc outside of the ring \citep{cazzoletti2018}. Simulations by \citet{lobo2015} show that after a vortex forms at the edge of a planet-induced gap, the surface density can be enhanced further out in the disk than at the initial vortex location. Such density enhancement triggers again the Rossby wave instability and a second generation vortex may form beyond the primary, leaving only the outermost vortex once the first one is damped.
Thus, a second-generation vortex may explain the location of the observed arc in the outer disk of HD\,143006.

We note that the vortex-like structure is not associated to spiral arms in scattered light \citep{benisty2018}, unlike the well-studied cases of MWC\,758 \citep{dong2018b}, HD\,142527 \citep{avenhaus2014} and HD~135344B \citep{stolker2016}, which in addition are classified as transition disks from their spectral energy distribution and whose central stars are Herbig ABe objects.

\subsection{Substructures seen in scattered light}
\label{sec:SPHERE}

Figure~\ref{fig:SPHERE} presents  our continuum image (left), the polarized scattered light coronagraphic image of HD\,143006 obtained with VLT/SPHERE\footnote{We note that the SPHERE infrared (1.2\,$\mu$m) image has a resolution of $37\times$37\,mas, comparable to the spatial resolution of the ALMA observations \citep{benisty2018}.} (middle), and the peak intensity of the CO line at each velocity, also known as moment 8 map (right). The scattered light image is a good tracer of small micron-sized dust grains located in tenuous disk surface layers, while the ALMA continuum image traces the millimeter-sized dust grains at the midplane. The peak intensity of the (optically thick) gas emission from CO (moment 8 image) also traces the upper disk layers. 

Each pixel of the scattered-light image (middle panel, Figure\,\ref{fig:SPHERE}) was scaled by $r^{2}$, where $r$ is the distance to the central star, to compensate for the drop off in stellar illumination and allow a better detection of faint outer disk features. In this image, a broad shadow in the West, covering half of the disk, is present. In addition, the SPHERE image shows from inside out: a  gap/cavity beyond the coronagraph radius ($\sim$80\,mas, 13\,au), a non-symmetric ring with two narrow shadow lanes aligned along the North-South direction, a gap not completely devoid of emission, and a non-symmetric outer disk. Both the outer disk and the inner ring present an over-brightness along a small range of position angles (PA$\sim$100-170$^\circ$, see contours in Figure~\ref{fig:SPHERE}).

\subsubsection{Comparison with CO emission}

As the CO line is optically thick, its intensity probes the temperature of the emitting gas, which, in turn, depends on the amount of starlight received at the disk surface. 
In scattered light, the outer part of the disk is not detected in the West, indicating a drop of irradiation. To first approximation, the temperature of the CO emitting layer should scale as the received luminosity to the $1/4$ power, thus, we should also expect an East-West brightness asymmetry in the peak intensity CO map (left panel, Figure~\ref{fig:SPHERE}).
We measure a East-West contrast in the CO peak intensity map of roughly $\sim1.2$.
The difference between the level of asymmetry in the scattered-light image and in the moment 8 map is likely related to the different depths of the $\tau\sim$1 layers in the two tracers.  We note however that no substructure in the CO channel maps or moment maps is found to coincide with the narrow shadow lanes in the inner ring of the scattered light image.

The absence of an over-brightness in the CO peak intensity map at the location of the millimeter arc suggests that the over-brightness observed in scattered light does not only originate from an effect of a stronger irradiation. And since the continuum emission in the arc is not optically thick ($\tau \sim 0.4$ at the peak) this is unlikely an issue of continuum subtraction of optically thick emission on the CO line.
As the inner disk is not probed by the SPHERE data, we cannot further compare the observed depletion of the CO column density inside of $\sim$15\,au ($\sim90$\,mas) discussed in \S4.1. Interestingly, the ``kink'' seen in the channel maps (\S3.2, also see Figure~\ref{fig:kink} in the appendix) roughly coincides in radius and azimuthal extent with the over-brightness seen in scattered light along the inner ring.

\subsubsection{Comparison with dust continuum}
\begin{figure*}[!t]
\begin{center}
\includegraphics[width=0.9\textwidth]{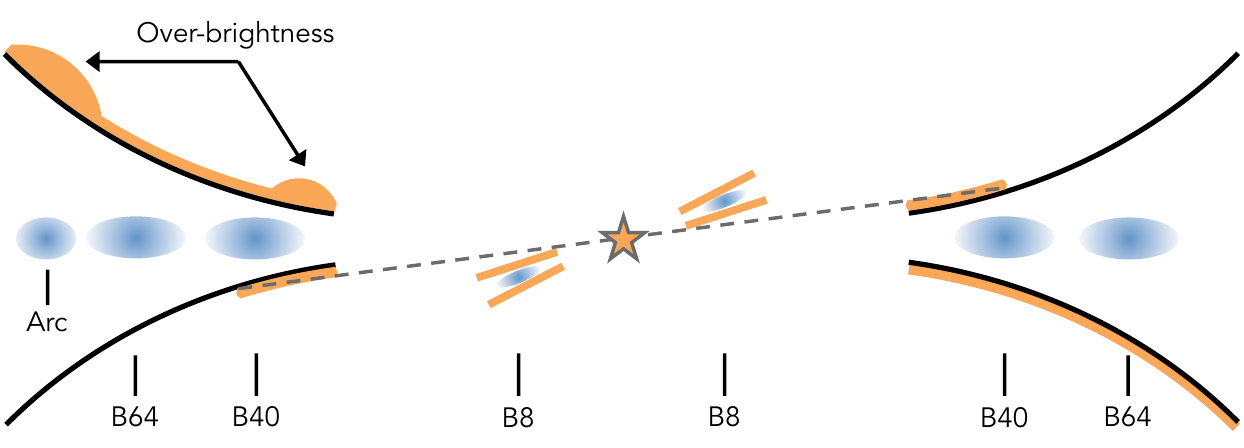}
\end{center}
\caption{Schematics of the HD\,143006 disk, comparing features traced by ALMA (detected rings/arc in color blue) and  scattered-light SPHERE observations (illuminated surface in orange). The innermost ring detected by ALMA (namely B8) has an inner cavity and it is misaligned with respect to the outer disk, explaining the asymmetric illumination observed on the scattered-light images by \citet{benisty2018}, where one side of the disk is directly illuminated (top-left and bottom-right surface of the disk in the figure) while the other side is partially illuminated (inner top-right and bottom-left surface). A gap in scattered-light emission is observed inwards of B40, while between B8 and B40 there is little millimeter emission observed. This is indicative of both small and large grains depletion the inner and outer disks. Two over-brightness are seen in scattered-light, which line-up azimuthally with the arc seen at ALMA wavelengths. The outermost over-brightness appears radially collocated with the arc while the innermost feature has no counterpart in the ALMA continuum image.
\label{fig:cartoon}}
\end{figure*}

The comparison between the ALMA continuum and the SPHERE scattered light images indicates striking differences. Apart from the bright prominent arc in the outer disk, none of the features appear co-radial and none of the azimuthal asymmetries seen in scattered light have counterparts in the continuum image. Such differences are expected, as the two images trace distinct layers of the disk and different dust particles. The scattered light image shows the regions of the disk that are directly lit by the star and its appearance strongly depends on the shape of the disk surface, while the 1.25\,mm continuum image shows the midplane features, whose brightness depend on the dust density, temperature, and opacity. An schematic of the features observed in the HD\,143006 disk is presented in Figure\,\ref{fig:cartoon} and will be discussed here. \\

\indent \textbf{Radial distribution.} While B8 is masked by the SPHERE coronagraph, the inner ring in scattered light appears located inside B40 (see middle panel of Figure~\ref{fig:SPHERE}), indicating that micron-sized grains are extending further in than the mm-sized grains. Such a spatial segregation by particle size is a natural outcome of dust trapping by a massive planet \citep{rice2006,zhu2012,dejuanovelar2013,pinilla2015} and has already been observed in transition disks \citep[e.g.,][]{garufi2013}. If the inner scattered-light ring traces the edge of the gap, it would be directly illuminated by the star and ``puff-up''. We propose that B40 lies in its shadow, which is supported by a darker region seen just beyond the scattered-light inner ring (on the East side of the disk), and that the outer disk re-emerges from the shadow at a larger distance from the star \citep[e.g.,][]{dullemond2004,isella2018}. 

We note that a projected radial offset between the features seen in scattered light and in the millimeter could be expected due to the disk inclination and position angle, the opening angle of the scattering surface, and the vertical structure of the ring \citep{dong2018b}. However, such an effect is considerable only for high disk inclinations. In that case, on the near side of the disk the scattered-light ring should appear inside B40, while on the far side, it should appear outside of it. That is not what we observe: the inner scattered-light ring is inside B40 at all position angles at which it is detected. 

It is possible that a companion between B8 and B40 is shaping the disk, leading to the observed radial segregation and to a dust-depleted gap in both tracers of small and large dust grains. 
In this paper series, \citet{Shangjia2018} constrain a planet mass of $\sim 10-20 M_{Jup}$ based solely on the deep gap observed in ALMA images between B8 and B40. Such a high mass companion is consistent with the hydrodynamic simulations coupled with dust evolution from \citet{dejuanovelar2013}, which require a planet more massive than $9M_{Jup}$ to be responsible for the observed radial segregation by particle size.
Although we find a strong depletion in small grains that are well coupled to the gas, we do not detect clear depletion in CO emission between B8 and B40, suggesting that dust grains are filtered and that gas can still flow through the gap, similar to what is often seen in transition disks \citep{pinilla2015b}.

\indent \textbf{Azimuthal asymmetries.}  The large East/West shadow (and the additional two narrow shadow lanes) observed in the SPHERE image can be explained by a moderate misalignment of $\sim$20-30$^\circ$ between the inner and outer disk  \citep[][see also schematics in Figure~\ref{fig:cartoon}]{benisty2018}. As shown in the residual map of Figure~\ref{fig:modelCont}, B40 and B64 are  symmetric, and we find no evidence for a counterpart of the East/West shadow in the continuum observations that would indicate inhomogeneities in the dust temperature. This supports the finding of \citet{rings}, where the rings in HD\,143006 are optically thin. Indeed, a configuration in which the inner and outer disks are moderately misaligned (see Figure~\ref{fig:cartoon}) would lead to e.g. the West side of the disk in the shadow while the East side is irradiated, for the front side of the disk that is facing us. On the backward-facing side of the disk (not seen by the observer) the opposite happens: the East side is in the shadow while the West is irradiated.
If the disk is vertically optically thin, and the continuum emission traces the full vertical extent of the disk, the asymmetry will cancel out between the two sides, and the continuum emission would appear symmetric. We note however the presence of residuals (up to 7$\sigma$) in the south (PA$\sim$180$^\circ$) at the location where the East/West shadow starts.  As in the CO data, there is no evidence for counterparts to the narrow shadow lanes in the continuum observations, unlike what is observed in two transition disks so far \citep[DoAr\,44, J1604;][]{casassus2018,pinilla2018}. 

The bright arc in the ALMA image coincides with the over-brightness seen in the outer disk of the SPHERE image (see contours in Figure\,\ref{fig:SPHERE} and diagram in Figure~\ref{fig:cartoon}). If the arc traces an over-density in large grains, we also expect an enhancement of small particles due to  fragmentation, and if turbulent mixing over the vertical height of the disk allows it, a small-particle enhancement should be seen from the surface layers as well. Additionally, the small grains seem to extend further, and over a wider range of position angles, than the larger particles, which is one of the predictions of dust trapping in a vortex \citep{baruteau2016}. Future observations at shorter/longer ALMA wavelengths can test this idea by measuring the spread/concentration of particles over the arc location.

The inner ring from the scattered-light image shows a strong over-brightness that has no counterpart in the ALMA continuum image (see contours inside B40 in Figure\,\ref{fig:SPHERE}), and that is likely due to an irradiation effect and local change in scale height. Interestingly, it lies over a range of position angles similar to the one over which the arc extends in the ALMA continuum image. If the line-up is not coincidental, it must be related to a radiative effect as any physical structure at such different radii would shear away the alignment quickly due to different angular rotation velocities. 
However, it is possible that the over-brightness along the first scattered-light ring further extends to the West (in the shadowed region), and that it covers a much broader range of position angles. In that case, the line-up could be coincidental.  

In general, such differences likely mean that the small grains in the surface layers  are only marginally affected by dynamical processes (while large grains in the midplane are), but are instead very much affected by irradiation processes.

\subsection{A misalignment between inner and outer disk}
Our continuum emission modeling indicates that the inner and outer disk do not share the same geometry, which is expected based on the shadows observed in the SPHERE image \citep{benisty2018}. The inner disk appears inclined by 24$^\circ$ with a PA of 164$^\circ$, while the outer disk has a lower inclination of 17$^\circ$ with a PA of 176$^\circ$, this difference in geometry is of high statistical significance (see Table \ref{tab:dustmodel}). Based on these values, we can estimate the misalignment angle between the inner and outer disks, defined as the angle between the normal vectors to the disks. Assuming the values above, we find a small misalignment of 8$^\circ$. However, this assumes that the inner and outer disks share the same near side of the disk (the side closer to us), which cannot be determined with the continuum data alone. If instead, the near sides do not coincide (e.g., the near side of the outer disk is in the East, while the near side of the inner disk is in the West), the misalignment would be much larger, of 41$^\circ$ (computed using $i=-24^\circ$ for the inner disk, the rest of the parameters being the same as above). In this case, once projected onto the plane of the sky, the inner disk rotation would appear as counter-rotating with respect to the outer disk. This is not something seen in the CO data (at an angular resolution of 66$\times$49\,mas, $\sim9$au), in particular at the highest velocities. Nevertheless, the fit of the intensity-weighted velocity map with a Keplerian disk model shows residuals inside $0.15''$, supporting a different geometry than the one of the outer disk. 

The location of the shadows seen in scattered light depend on the inner and outer disk geometry, and on the height of the scattering surface ($z_{\rm{scat}}$) of the disk where the shadows are seen. Using the equations developed in \citet{min2017}, we find a $z_{\rm{scat}}/R$ of $\sim$0.03 and $\sim$0.16 at $\sim$18\,au (the location of the inner scattered light ring), for misalignment angles of $7\degree$ and $41\degree$, respectively.  Since the pressure scale height of the disk, $H_p$, should be smaller than the scattering surface (by a factor of $\sim$2-4), we expect $H_p/R<0.015$ and $<$0.08 at $\sim$18\,au, for misalignment angles of $7\degree$ and $41\degree$, respectively. From the standard temperature profile assumed in \citet{rings}, we estimate that at 18\,au $H_p/R\sim0.04$. Given the small $H_p/R$ expected for a $7\degree$ misalignment and the larger value of the pressure scale height as estimated above from standard assumptions, we favor the larger misalignment value of $41\degree$.

The overall morphology of the shadows in the scattered light image were reproduced by  a circumbinary disk that is broken and misaligned by $\sim$30$^\circ$ \citep{benisty2018}, due to an inclined equal-mass binary \citep[see the hydrodynamical simulations by][]{facchini2018}. We note however, that the value of the misalignment needed to reproduce the scattered light image is model-dependent, and depends on the exact geometry of the inner disk as well as on the shape of the outer disk rim considered in the model.

The presence of an equal mass binary companion, as well as of any companion with a mass ratio larger than q=0.2 (corresponding to 0.3\,M$_{\odot}$), can be ruled out as it  would have been detected by imaging and interferometric surveys \citep{kraus2008,benisty2018}. However, an inclined low-mass stellar companion, as the one detected in the wide gap of HD\,142527 (with a mass $\sim$0.13\,M$_{\odot}$), could be responsible for the misalignment \citep{price2018}, and secular precession resonances can result in large misalignments for companions with mass ratio of 0.01-0.1 \citep{owen2017}.  A massive planet could in principle also lead to a misaligned inner disk as long as its angular momentum is larger than that of the inner disk \citep[e.g.][]{xg2013, bitsch2013, matsakos2017}, i.e., in cases where the inner disk is depleted. 

In any case, it is not clear whether the putative companion responsible for the misalignment in HD\,143006 should be located in the gap between B8 and B40, or between B8 and the dust sublimation edge. The spectral energy distribution indicates the presence of hot dust close to the sublimation radius, which was spatially resolved by near-infrared interferometry \citep{lazareff2017}, suggestive of the presence of (at least) another ring of small dust in the innermost au where mm-sized grains are depleted.

\section{Conclusions} \label{sec:conclusions}

From the DSHARP ALMA observations of the HD\,143006 protoplanetary disk that reach $\sim7$au in spatial resolution, and its comparison with existing scattered-light observations at similar spatial resolution, we conclude the following: 

\begin{itemize}
 
\item In terms of substructure, the dust continuum emission from HD\,143006 reveals three bright rings, two dark gaps, and an arc at the edge of the dusty disk. The CO observations at similar angular resolution exhibit a depletion of gas emission in the inner disk with no significant features, except for a deviation or ``kink'' from the Keplerian rotation pattern over a few red-shifted channels close to the systemic velocity.

\item From different tracers of the disk structure we find further evidence for a misalignment between the inner and outer disk: a fit to the disk Keplerian velocities with a global/single disk geometry does not account well for the inner disk kinematics, while modeling of dust continuum emission results in a more inclined inner disk (as traced by B8) than the outer disk (as traced by B40, B64, and the south-east arc). These findings are in agreement with existing VLT/SPHERE images that suggest a disk misalignment from the presence of shadows.

\item The prominent south-east arc in the ALMA 1.25\,mm image resolves into three peaks along its azimuthal extent. The counterpart to this arc in the scattered-light image shows a broader radial and azimuthal extent, indicative of segregation by particle size as would be expected for a dust-trapping vortex at this location. Future observations at longer wavelengths should be able to test this scenario.

\item The bright rings have increasingly larger radial widths with increasing distance from the star. We find evidence for radial segregation by particle size at the outer edge of the gap between B8 and B40, and a strong depletion of small and large grains in the gap. These are consistent with a companion carving the gap. However, no dark annulus is observed in CO emission at this radius, suggesting that dust grains are filtered but gas can still flow through the gap. 
 
\end{itemize}

Future observations at longer millimeter wavelengths will allow us to determine if there is efficient trapping of dust at the substructures location, if these are due to localized pressure maxima, and in particular to understand if the arc traces a vortex. Observing gas tracers at high spectral resolution will also be  fundamental to elucidate the absolute misalignment of the inner disk and to determine the velocity structure at the arc location. 

\acknowledgments
We are thankful to S.~Facchini, A.~Juh\'asz and R.~Teague for insightful discussions. 
This paper makes use of ALMA data \dataset[ADS/JAO.ALMA\#2016.1.00484.L]{https://almascience.nrao.edu/aq/?project\_code=2016.1.00484.L}  and \dataset[ADS/JAO.ALMA\#2015.1.00964.S]{https://almascience.nrao.edu/aq/?project\_code=2015.1.00964.S}). ALMA is a partnership of ESO (representing its member states), NSF (USA) and NINS (Japan), together with NRC (Canada), MOST and ASIAA (Taiwan), and KASI (Republic of Korea), in cooperation with the Republic of Chile. The Joint ALMA Observatory is operated by ESO, AUI/NRAO and NAOJ. 
L.P. acknowledges support from CONICYT project Basal AFB-170002 and from FCFM/U. de Chile Fondo de Instalaci\'on Acad\'emica. M.B. acknowledges funding from ANR of France under contract number ANR-16-CE31-0013 (Planet Forming disks). 
S. A. and J. H. acknowledge funding support from NASA Exoplanets Research Program grant 80NSSC18K0438. 
T.B. acknowledges funding from the European Research Council (ERC) under the European Union’s Horizon 2020 research and innovation programme under grant agreement No 714769. 
C.P.D. acknowledges support by the German Science Foundation (DFG) Research Unit FOR 2634, grants DU 414/22-1 and DU 414/23-1.
V.V.G. and J.C acknowledge support from the National Aeronautics and Space Administration under grant No. 15XRP15\_20140 issued through the Exoplanets Research Program. 
J.H. acknowledges support from the National Science Foundation Graduate Research Fellowship under Grant No. DGE-1144152. 
A.I. acknowledges support from the National Aeronautics and Space Administration under grant No. NNX15AB06G issued through the Origins of Solar Systems program, and from the National Science Foundation under grant No. AST-1715719.
L. R. acknowledges support from the ngVLA Community Studies program, coordinated by the National Radio Astronomy Observatory, which is a facility of the National Science Foundation operated under cooperative agreement by Associated Universities, Inc. 
Z. Z. and S. Z.acknowledges support from the National Aeronautics and Space Administration through the Astrophysics Theory Program with Grant No. NNX17AK40G and Sloan Research Fellowship. Simulations are carried out with the support from the Texas Advanced Computing Center (TACC) at The University of Texas at Austin through XSEDE grant TG-AST130002.

\appendix

\section{Additional Figures}

\begin{figure*}[!ht]
\includegraphics[width=\textwidth]{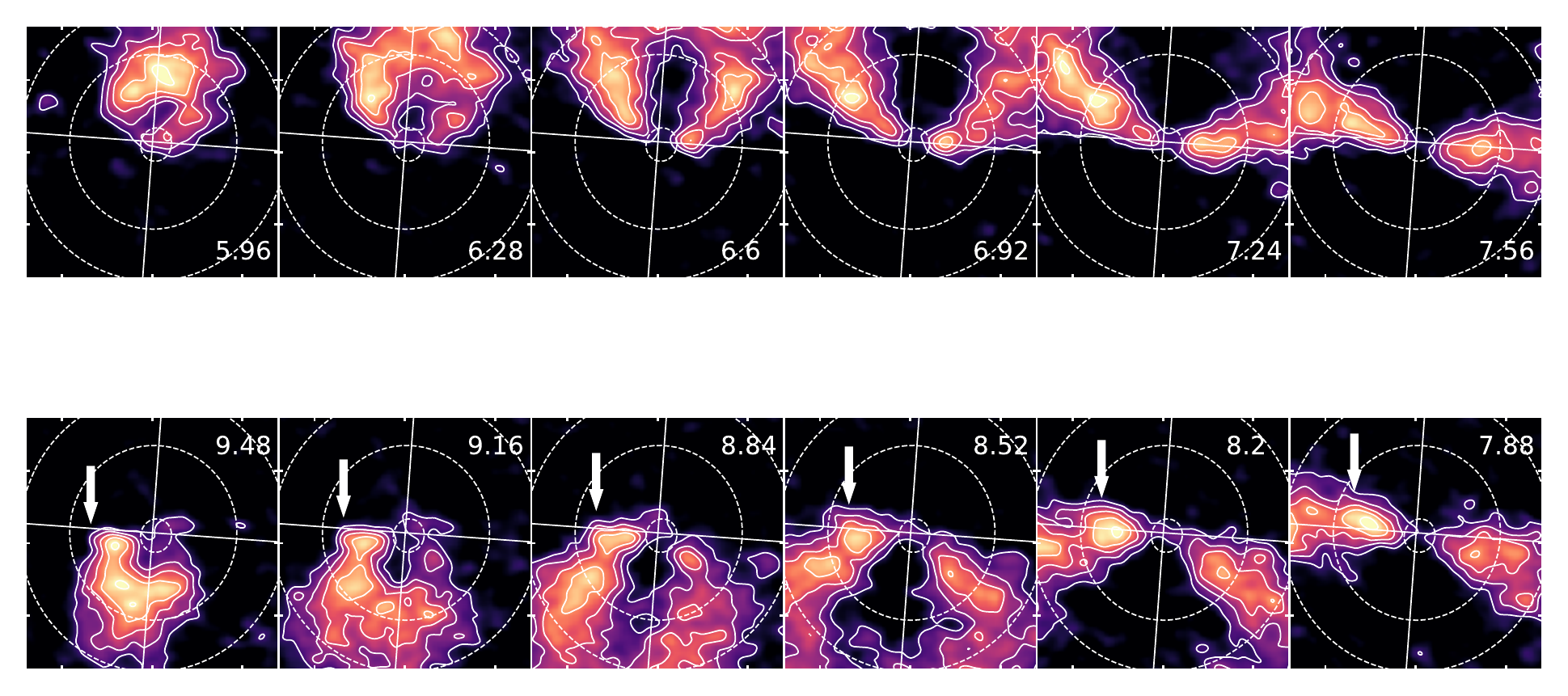}
\caption{
Zoom-in to the channel maps near the systemic velocity of HD\,143006, with blue-shifted and red-shifted channels on the top and bottom, respectively. Note that the red-shifted channel ordering has been flipped (velocity increasing towards the left) to aid the comparison with its symmetric channels on the blue-shifted side of the line. The arrow marks the perturbation or ``kink'' seen at similar PA as the bright over-density in scattered light. Notice also how different the emission is in the region marked by the arrow compared to the symmetric channel on the top.
\label{fig:kink}}
\end{figure*}

\begin{figure*}[ht!]
\includegraphics[width=\textwidth]{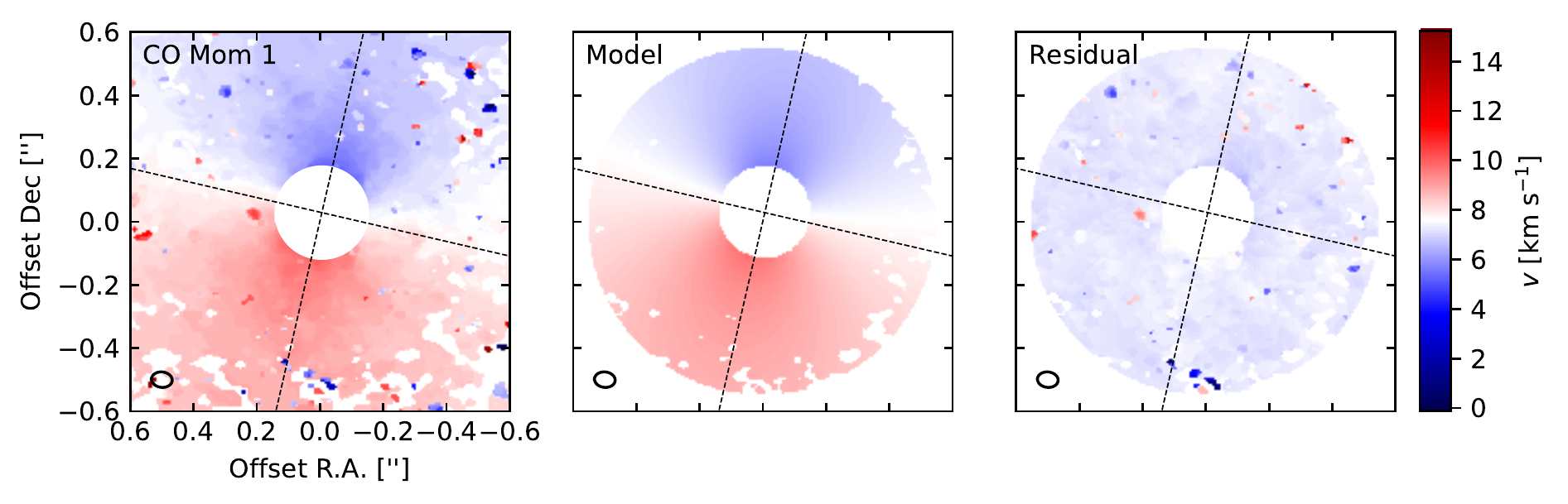}
\includegraphics[width=\textwidth]{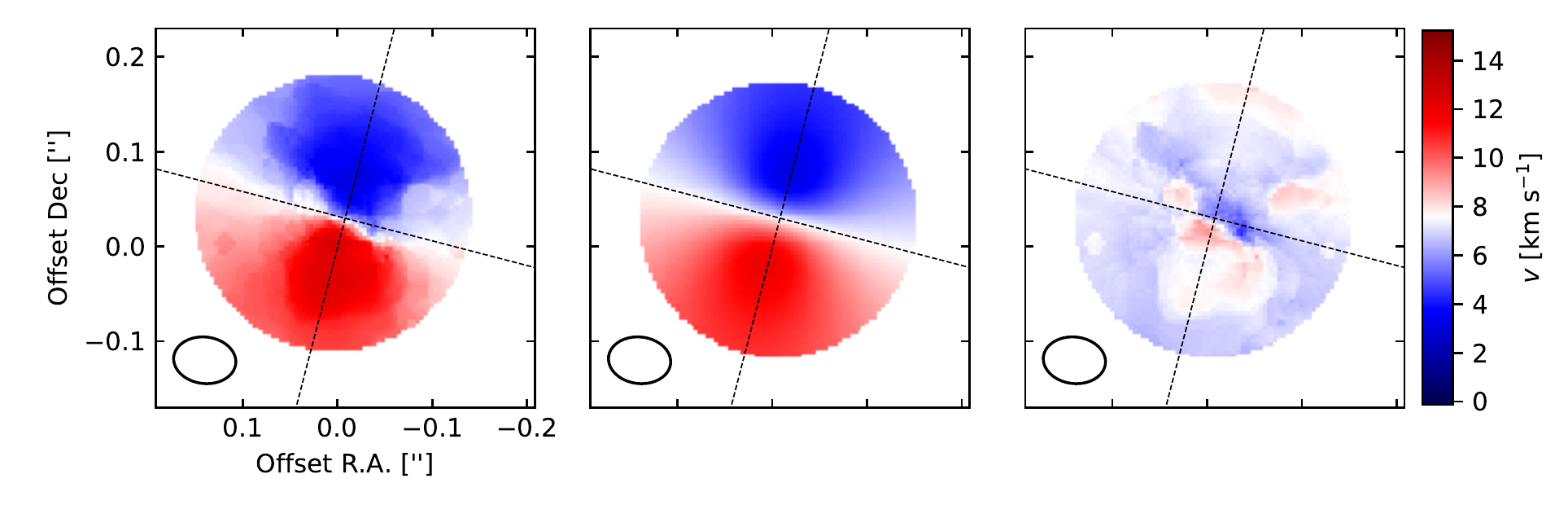}
\caption{
Comparison between the intensity-weighted velocity field around HD\,143006 (right panels), the best-fit model for a Keplerian razor-thin disk (middle panels), and the residual velocity field map obtained from subtracting the best-fit model from the observations (left panels).
The top panels show the fit results when the inner disk is masked (inside of 0.15$''$), while the bottom panels have been zoomed in and show the fit results when the outer disk is masked (outside of 0.15$''$).
The PA of the disk that is obtained from each fit ($167.0\degree$ for the top panels, $165.6\degree$ for the bottom panels), are indicated by dashed lines. \label{fig:restrictedmom1}}
\end{figure*}

Here we present additional supporting figures. First, a zoom-in to the channel maps near the systemic velocity that show a ``kink'' on the red-shifted channels (bottom panels of Figure \ref{fig:kink}) that is not present on the blue-shifted channels (top panels of Figure \ref{fig:kink}). Second, Figure~\ref{fig:restrictedmom1} shows how the Keplerian velocity fit differs when the intensity-weighted velocity field map has been masked inside 0.15$''$ (top panels, Figure~\ref{fig:restrictedmom1}) and masked outside of $0.15''$ (bottom panels, Figure~\ref{fig:restrictedmom1}).

\begin{thebibliography}{}

\bibitem[ALMA Partnership et al.(2015)]{ALMA} ALMA Partnership, Brogan, C.~L., P{\'e}rez, L.~M., et al.\ 2015, \apj, 808, L3.
\bibitem[Andrews et al.(2018)]{survey} Andrews, S.~M., Huang, J., P\'erez, L.~M., et al.\ 2018, \apj, in press. 
\bibitem[Ataiee et al.(2013)]{ataiee2013} Ataiee, S., Pinilla, P., Zsom, A., et al.\ 2013, \aap, 553, L3.
\bibitem[Avenhaus et al.(2014)]{avenhaus2014} Avenhaus, H., Quanz, S.~P., Schmid, H.~M., et al.\ 2014, \apj, 781, 87.
\bibitem[Avenhaus et al.(2018)]{avenhaus2018} Avenhaus, H., Quanz, S.~P., Garufi, A., et al.\ 2018, \apj, 863, 44.
\bibitem[Bai \& Stone(2014)]{bai2014} Bai, X.-N., \& Stone, J.~M.\ 2014, \apj, 796, 31 
\bibitem[Barenfeld et al.(2016)]{barenfeld2016} Barenfeld, S.~A., Carpenter, J.~M., Ricci, L., et al.\ 2016, \apj, 827, 142.
\bibitem[Barge \& Sommeria(1995)]{barge1995} Barge, P., \& Sommeria, J.\ 1995, \aap, 295, L1 
\bibitem[Baruteau \& Zhu(2016)]{baruteau2016} Baruteau, C., \& Zhu, Z.\ 2016, \mnras, 458, 3927 
\bibitem[Benisty et al.(2017)]{benisty2017} Benisty, M., Stolker, T., Pohl, A., et al.\ 2017, \aap, 597, A42.
\bibitem[Benisty et al.(2018)]{benisty2018} Benisty, M., Juh{\'a}sz, A., Facchini, S., et al.\ 2018, \aap, 619, A171 
\bibitem[B{\'e}thune et al.(2017)]{bethune2017} B{\'e}thune, W., Lesur, G., \& Ferreira, J.\ 2017, \aap, 600, A75 
\bibitem[Birnstiel et al.(2010)]{birnstiel2010} Birnstiel, T., Dullemond, C.~P., \& Brauer, F.\ 2010, \aap, 513, A79.
\bibitem[Birnstiel et al.(2013)]{birnstiel2013} Birnstiel, T., Dullemond, C.~P., \& Pinilla, P.\ 2013, \aap, 550, L8.
\bibitem[Bitsch et al.(2013)]{bitsch2013} Bitsch, B., Crida, A., Libert, A.-S., et al.\ 2013, \aap, 555, A124.
\bibitem[Bouvier et al.(2007)]{bouvier2007} Bouvier, J., Alencar, S.~H.~P., Boutelier, T., et al.\ 2007, \aap, 463, 1017.
\bibitem[Brauer et al.(2007)]{brauer2007} Brauer, F., Dullemond, C.~P., Johansen, A., et al.\ 2007, \aap, 469, 1169.
\bibitem[Casassus et al.(2013)]{casassus2013} Casassus, S., van der Plas, G., M, S.~P., et al.\ 2013, \nat, 493, 191.
\bibitem[Casassus et al.(2015)]{casassus2015} Casassus, S., Marino, S., P{\'e}rez, S., et al.\ 2015, \apj, 811, 92.
\bibitem[Casassus et al.(2018)]{casassus2018} Casassus, S., Avenhaus, H., P{\'e}rez, S., et al.\ 2018, \mnras, 477, 5104.
\bibitem[Cazzoletti et al.(2018)]{cazzoletti2018} Cazzoletti, P., van Dishoeck, E.~F., Pinilla, P., et al.\ 2018, \aap, 619, A161 
\bibitem[Cody et al.(2014)]{cody2014} Cody, A.~M., Stauffer, J., Baglin, A., et al.\ 2014, \aj, 147, 82.
\bibitem[Crida \& Morbidelli(2007)]{crida2007} Crida, A., \& Morbidelli, A.\ 2007, \mnras, 377, 1324.
\bibitem[de Juan Ovelar et al.(2013)]{dejuanovelar2013} de Juan Ovelar, M., Min, M., Dominik, C., et al.\ 2013, \aap, 560, A111.
\bibitem[Dittrich et al.(2013)]{dittrich2013} Dittrich, K., Klahr, H., \& Johansen, A.\ 2013, \apj, 763, 117 
\bibitem[Dong et al.(2018)]{dong2018b} Dong, R., Liu, S.-. yuan ., Eisner, J., et al.\ 2018, \apj, 860, 124.
\bibitem[Dullemond \& Dominik(2004)]{dullemond2004} Dullemond, C.~P., \& Dominik, C.\ 2004, \aap, 417, 159 
\bibitem[Dullemond et al.(2018)]{dullemond2018} Dullemond, C.-P., et al.\ 2018, \apj, in press. 
\bibitem[Facchini et al.(2018)]{facchini2018} Facchini, S., Juh{\'a}sz, A., \& Lodato, G.\ 2018, \mnras, 473, 4459.
\bibitem[Flock et al.(2017)]{flock2017} Flock, M., Fromang, S., Turner, N.~J., et al.\ 2017, \apj, 835, 230.
\bibitem[Foreman-Mackey et al.(2013)]{emcee} Foreman-Mackey, D., Hogg, D.~W., Lang, D., \& Goodman, J.\ 2013, \pasp, 125, 306 
\bibitem[Fu et al.(2014)]{fu2014} Fu, W., Li, H., Lubow, S., Li, S., \& Liang, E.\ 2014, \apjl, 795, L39 
\bibitem[Gaia Collaboration et al.(2018)]{Gaia2018} Gaia Collaboration, Brown, A.~G.~A., Vallenari, A., et al.\ 2018, \aap, 616, A1 
\bibitem[Garufi et al.(2013)]{garufi2013} Garufi, A., Quanz, S.~P., Avenhaus, H., et al.\ 2013, \aap, 560, A105.
\bibitem[Garufi et al.(2018)]{garufi2018} Garufi, A., Benisty, M., Pinilla, P., et al.\ 2018, \aap, 620, A94 
\bibitem[Huang et al.(2018a)]{rings} Huang, J., Andrews, S.~M., Dullemond, C.~P., et al.\ 2018, \apj, in press.
\bibitem[Huang et al.(2018b)]{spirals} Huang, J., Andrews, S.~M., P\'erez, L.~M., et al.\ 2018, \apj, submitted.
\bibitem[Isella et al.(2013)]{isella2013} Isella, A., P{\'e}rez, L.~M., Carpenter, J.~M., et al.\ 2013, \apj, 775, 30.
\bibitem[Isella \& Turner(2018)]{isella2018} Isella A., \& Turner, N.~J.\ 2018, \apj, 860, 27.
\bibitem[Isella et al.(2018)]{isellaLP} Isella, A., Huang, J., Andrews, S.~M., et al.\ 2018, \apj, in press. 
\bibitem[Johansen et al.(2009)]{johansen2009} Johansen, A., Youdin, A., \& Klahr, H.\ 2009, \apj, 697, 1269 
\bibitem[Klahr \& Henning(1997)]{klahr1997} Klahr, H.~H., \& Henning, T.\ 1997, \icarus, 128, 213 
\bibitem[Kraus et al.(2008)]{kraus2008} Kraus, A.~L., Ireland, M.~J., Martinache, F., et al.\ 2008, \apj, 679, 762.
\bibitem[Kretke, \& Lin(2007)]{kretke2007} Kretke, K.~A., \& Lin, D.~N.~C.\ 2007, \apj, 664, L55.
\bibitem[Lazareff et al.(2017)]{lazareff2017} Lazareff, B., Berger, J.-P., Kluska, J., et al.\ 2017, \aap, 599, A85 
\bibitem[Li et al.(2000)]{li2000} Li, H., Finn, J.~M., Lovelace, R.~V.~E., et al.\ 2000, \apj, 533, 1023.
\bibitem[Lobo Gomes et al.(2015)]{lobo2015} Lobo Gomes, A., Klahr, H., Uribe, A.~L., et al.\ 2015, \apj, 810, 94.
\bibitem[Lyra \& Lin(2013)]{Lyra} Lyra, W., \& Lin, M.-K.\ 2013, \apj, 775, 17 
\bibitem[Loomis et al.(2017)]{loomis2017} Loomis, R.~A., {\"O}berg, K.~I., Andrews, S.~M., et al.\ 2017, \apj, 840, 23.
\bibitem[Marino et al.(2015)]{marino2015} Marino, S., Perez, S., \& Casassus, S.\ 2015, \apj, 798, L44.
\bibitem[Matsakos \& K{\"o}nigl(2017)]{matsakos2017} Matsakos, T., \& K{\"o}nigl, A.\ 2017, \aj, 153, 60.
\bibitem[Meheut et al.(2012)]{meheut2012} Meheut, H., Meliani, Z., Varniere, P., et al.\ 2012, \aap, 545, A134.
\bibitem[Min et al.(2017)]{min2017} Min, M., Stolker, T., Dominik, C., et al.\ 2017, \aap, 604, L10.
\bibitem[Okuzumi et al.(2016)]{okuzumi2016} Okuzumi, S., Momose, M., Sirono, S.-. iti ., et al.\ 2016, \apj, 821, 82.
\bibitem[Owen, \& Lai(2017)]{owen2017} Owen, J.~E., \& Lai, D.\ 2017, \mnras, 469, 2834.
\bibitem[Paardekooper \& Mellema(2004)]{Paardekooper2004} Paardekooper, S.-J., \& Mellema, G.\ 2004, \aap, 425, L9 
\bibitem[Pecaut et al.(2012)]{pecaut2012} Pecaut, M.~J., Mamajek, E.~E., \& Bubar, E.~J.\ 2012, \apj, 746, 154 
\bibitem[P{\'e}rez et al.(2014)]{Perez2014} P{\'e}rez, L.~M., Isella, A., Carpenter, J.~M., \& Chandler, C.~J.\ 2014, \apjl, 783, L13 
\bibitem[Perez et al.(2015)]{seba2015} Perez, S., Dunhill, A., Casassus, S., et al.\ 2015, \apjl, 811, L5 
\bibitem[Pinilla et al.(2012a)]{pinilla2012a} Pinilla, P., Birnstiel, T., Ricci, L., et al.\ 2012, \aap, 538, A114.
\bibitem[Pinilla et al.(2012b)]{pinilla2012b} Pinilla, P., Benisty, M., \& Birnstiel, T.\ 2012, \aap, 545, A81.
\bibitem[Pinilla et al.(2015)]{pinilla2015} Pinilla, P., de Juan Ovelar, M., Ataiee, S., et al.\ 2015, \aap, 573, A9.
\bibitem[Pinilla et al.(2015)]{pinilla2015b} Pinilla, P., de Boer, J., Benisty, M., et al.\ 2015, \aap, 584, L4 
\bibitem[Pinilla et al.(2018)]{pinilla2018} Pinilla, P., Benisty, M., de Boer, J., et al.\ 2018, \apj, 868, 85 
\bibitem[Pinte et al.(2018)]{pinte2018} Pinte, C., Price, D.~J., M{\'e}nard, F., et al.\ 2018, \apj, 860, L13.
\bibitem[Preibisch et al.(2002)]{preibisch2002} Preibisch, T., Brown, A.~G.~A., Bridges, T., Guenther, E., \& Zinnecker, H.\ 2002, \aj, 124, 404 
\bibitem[Price et al.(2018)]{price2018} Price, D.~J., Cuello, N., Pinte, C., et al.\ 2018, \mnras, 477, 1270.
\bibitem[Reg{\'a}ly et al.(2012)]{regaly2012} Reg{\'a}ly, Z., Juh{\'a}sz, A., S{\'a}ndor, Z., et al.\ 2012, \mnras, 419, 1701.
\bibitem[Reg{\'a}ly et al.(2017)]{regaly2017} Reg{\'a}ly, Z., Juh{\'a}sz, A., \& Neh{\'e}z, D.\ 2017, \apj, 851, 89.
\bibitem[Rice et al.(2006)]{rice2006} Rice, W.~K.~M., Armitage, P.~J., Wood, K., et al.\ 2006, \mnras, 373, 1619.
\bibitem[Rigliaco et al.(2015)]{rigliaco2015} Rigliaco, E., Pascucci, I., Duchene, G., et al.\ 2015, \apj, 801, 31.
\bibitem[Rosenfeld et al.(2012)]{rosenfeld2012} Rosenfeld, K.~A., Qi, C., Andrews, S.~M., et al.\ 2012, \apj, 757, 129.
\bibitem[Rosenfeld et al.(2013)]{rosenfeld2013} Rosenfeld, K.~A., Andrews, S.~M., Hughes, A.~M., Wilner, D.~J., \& Qi, C.\ 2013, \apj, 774, 16 
\bibitem[Rosenfeld et al.(2014)]{rosenfeld2014} Rosenfeld, K.~A., Chiang, E., \& Andrews, S.~M.\ 2014, \apj, 782, 62.
\bibitem[Simon \& Armitage(2014)]{simon2014} Simon, J.~B., \& Armitage, P.~J.\ 2014, \apj, 784, 15 
\bibitem[Stolker et al.(2016)]{stolker2016} Stolker, T., Dominik, C., Avenhaus, H., et al.\ 2016, \aap, 595, A113.
\bibitem[Suriano et al.(2018)]{suriano2018} Suriano, S.~S., Li, Z.-Y., Krasnopolsky, R., \& Shang, H.\ 2018, \mnras, 477, 1239 
\bibitem[Tazzari et al.(2018)]{Galario} Tazzari, M., Beaujean, F., \& Testi, L.\ 2018, \mnras, 476, 4527.
\bibitem[Teague et al.(2018)]{teague2018} Teague, R., Bae, J., Bergin, E.~A., et al.\ 2018, \apj, 860, L12.
\bibitem[van der Marel et al.(2015)]{Marel2015} van der Marel, N., van Dishoeck, E.~F., Bruderer, S., P{\'e}rez, L., \& Isella, A.\ 2015, \aap, 579, A106 
\bibitem[Walsh et al.(2017)]{walsh2017} Walsh, C., Daley, C., Facchini, S., et al.\ 2017, \aap, 607, A114.
\bibitem[Weaver et al.(2018)]{weaver2018} Weaver, E., Isella, A., \& Boehler, Y.\ 2018, \apj, 853, 113.
\bibitem[Xiang-Gruess \& Papaloizou(2013)]{xg2013} Xiang-Gruess, M., \& Papaloizou, J.~C.~B.\ 2013, \mnras, 431, 1320.
\bibitem[Zhang et al.(2018)]{Shangjia2018} Zhang, S., Zhu, Z., Huang, J., et al.\ 2018, \apj, in press. 
\bibitem[Zhang et al.(2015)]{zhang2015} Zhang, K., Blake, G.~A., \& Bergin, E.~A.\ 2015, \apjl, 806, L7 
\bibitem[Zhu et al.(2011)]{zhu2011} Zhu, Z., Nelson, R.~P., Hartmann, L., et al.\ 2011, \apj, 729, 47.
\bibitem[Zhu et al.(2012)]{zhu2012} Zhu, Z., Nelson, R.~P., Dong, R., Espaillat, C., \& Hartmann, L.\ 2012, \apj, 755, 6 
\bibitem[Zhu \& Stone(2014)]{zhu2014} Zhu, Z., \& Stone, J.~M.\ 2014, \apj, 795, 53.
\bibitem[Zhu \& Baruteau(2016)]{zhu2016} Zhu, Z., \& Baruteau, C.\ 2016, \mnras, 458, 3918.

\end{thebibliography}
\end{document}